\documentstyle[twocolumn,floats,aps,psfig]{revtex}
\begin{document}
\draft
\title{On the Functional Integral Theory of Systems with Kinematical Interaction}
\author{L.V. Popovich and M.V. Medvedev}
\address{Physics Department, University of California at San Diego, La Jolla, CA 92093}
\maketitle

\begin{abstract}
We propose a systematic
way to investigate the low-temperature thermodynamic properties of 
quantum spin systems subject to the restriction that only a finite 
number of bosons may occupy a single lattice site.
Such a kinematical interaction results in appearance of a temperature 
dependent chemical potential. Its low-temperature asymptotics is 
calculated self-consistently using the functional integration technique.
\end{abstract}
\pacs{75.10.-b, 05.50.+q}

To systematically calculate the thermodynamic properties of a two-dimensional 
(2D) quantum ferromagnet at low temperatures remains an unsolved problem of
the spin wave theory \cite{spin-wave,HP}. The main difficulty is in how to 
consistently calculate both dynamical and kinematical \cite{kinem} 
interactions of magnons. Such interactions are believed to vanish the
average spin (magnetization) of a system, as it naturally follows from 
the absence of long range order in 2D systems at low temperatures.

Recently, a few different approaches to this problem have been developed.
In his ``modified spin wave theory'',  Takahashi \cite{[2]} uses the 
Holstein-Primakoff representation to introduce (by hands) a chemical 
potential for a system of bosons. This chemical potential is essentially
nonzero in 2D systems, even in the absence of magnetic field. Thus,
the zero magnetization condition, $\langle S_z\rangle=0$, is enforced.
Despite good agreement with numerical calculations, such an approach
seems to be not ultimately self-consistent. In another approach by Arovas 
and Auerbach \cite{[1]}, a functional integral for a partition function is
constructed using the generalized bosonic $SU(N)$ representation of 
spin algebra. Being calculated using the $1/N$ expansion around the
mean field ($N=\infty$) saddle point, the results were then extended to the 
case of $N=2$ [i.e., $SU(2)$]. The validity of this procedure has not,
however, been justified.

In this letter, we propose a self-consistent method to calculate
the partition function of a system of bosons with kinematical interaction,
i.e., the number of bosons on a lattice should not exceed some number $L$.
One should note that in contrast to Takahashi \cite{[2]}, who constructed 
a theory of low-dimensional ferromagnet by means of introducing ``by hand'' 
a chemical potential for Holstein-Primakoff bosons to satisfy the condition
$\langle S_{z} \rangle = 0$, we develop a theory with the chemical
potential as a result of controllable approximations.
We write the partition function in the functional integral representation.
In the one-loop approximation, we then show that the system of kinematically
interacting magnons is equivalent to that of non-interacting magnons
with some non-zero, temperature dependent chemical potential. 
Its low-temperature behavior is finally calculated. One should comment that the
theory of the systems with kinematical interaction which limits the maximum number
of particles per site is, in fact, closely related to the {\em parastatistics}
theory \cite{para} with essentially similar ``exclusion'' principles. 

We demonstrate the method on a simple model with the Hamiltonian:
\begin{equation}
\label{1}
H=\sum_{\bf k}\omega_{\bf k}b^{+}_{\bf k}b_{\bf k}\ , \quad
\omega_{\bf k} = \frac{J{\bf k}^{2}a^{2}}{2}\ , \quad
[b_{\bf k},b^{+}_{\bf q}] = \delta_{\bf k q}
\end{equation}
on a square lattice, where $a$ is a lattice constant, $\beta=1/t$ is 
the inverse temperature, and $b$ and $b^+$ are the Bose operators. 

\begin{figure}
\psfig{file=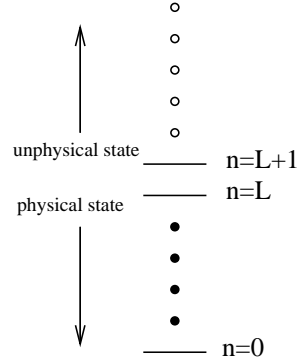,width=1.5in,angle=-90}
\caption{A lattice site; up to $L$ bosons may occupy this site.}
\label{fig1}
\end{figure}

As mentioned above, we consider the system of bosons with kinematical 
interaction. In other words, an each lattice site can be occupied by no 
more than $L$ bosons. Thus, we need a projecting operator to eliminate
all {\em unphysical states} with $n\ge L+1$ (see Fig.\ \ref{fig1}), 
while all {\em physical states} $0\le n\le L$ are to remain unchanged.
Heuristically, we define the projecting operator $P=\prod_{i} P_{i}$
as follows:
\begin{equation}
\label{3}
P_{i} = \ :e^{-b^{+}_{i}b_{i}}
\sum_{n=0}^{L}\frac{(b^{+}_{i}b_{i})^{n}}{n!}: \ ,
\end{equation}
where $:{\it O}:$ is normally ordered form of an operator ${\it O}$.
Here we used a useful identity: 
\begin{equation}
\label{4}
e^{\mu b^{+}b} = \ :e^{(e^{\mu}-1)b^{+}b}: 
\end{equation}
which can be proved straightforwardly.

Now the partition function of the system is:
\begin{equation}
\label{2}
Z = Sp\{P e^{-\beta H}\} .
\end{equation}
To obtain the functional representation for the partition function (\ref{2}),
it is instructive to transform to the coherent basis \cite{coherent}:
\begin{equation}
\label{5}
|u\rangle  = e^{-\frac{u^{*}u}{2}}
\sum_{n=0}^{\infty}\frac{u^{n}}{\sqrt{n!}}|n\rangle \ ,
\end{equation}
with the properties:
\begin{mathletters}
\begin{eqnarray}
\label{6}
&\int |u\rangle \langle u| \frac{du^{*}du}{\pi}
=\sum_{n=0}^{\infty}|n\rangle \langle n| = \hat{1} \ ,& \\
\label{7}
&b|u\rangle  = u|u\rangle  \ , \quad \langle u|b^{+} = \langle 
u|u^{*} \ , &\quad
\nonumber\\
&\langle v|u\rangle  = e^{-\frac{1}{2}(u^{*}u+v^{*}v) + v^{*}u} \ .&
\end{eqnarray}
\end{mathletters}
Using identity (\ref{6}), we rewrite the partition function as follows:
\begin{equation}
\label{8}
Z = \int \langle v|P|u\rangle \langle u|e^{-{\beta H}}|v\rangle  DuDv.
\end{equation}
Since all the operators are in normal ordered form, we use Eqs.\ (\ref{7}) 
to obtain the following matrix elements:
\begin{mathletters}
\begin{eqnarray}
\label{9}
\langle v_{i}|P_{i}|u_{i}\rangle
&=&e^{-\frac{1}{2}(u^{*}_{i}u_{i}+v^{*}_{i}v_{i})}
\sum_{n=0}^{L}\frac{(v^{*}_{i}u_{i})^{n}}{n!} \ , \\
\label{10}
\langle u|{e}^{-{\beta H}}|v\rangle
&=&e^{-\sum_{\bf k}(\frac{1}{2}(u^{*}_{\bf k}u_{\bf k}+
                             v^{*}_{\bf k}v_{\bf k})-
e^{-\beta \omega _{\bf k}}u^{*}_{\bf k}v_{\bf k})} \ .
\end{eqnarray}
\end{mathletters}
Thus, the exact partition function becomes:
\begin{equation}
\label{11}
Z = \int e^{-S(u,v)}DuDv,
\end{equation}
where the ``action'' $S(u,v)$ is
\begin{eqnarray}
\label{12}
S(u,v) &=& \sum_{\bf k}(u^{*}_{\bf k}u_{\bf k}+
                              v^{*}_{\bf k}v_{\bf k}-
e^{-\beta \omega _{\bf k}}v^{*}_{\bf k}u_{\bf k})
\nonumber\\
& &{ }-\sum_{i}\ln\Big(\sum_{n=0}^{L}\frac{(v^{*}_{i}u_{i})^{n}}{n!}\Big)
\end{eqnarray}
In the mean field approximation, we expand the logarithm up to the first 
nonlinear term:
\begin{eqnarray}
\label{13}
\ln\sum_{n=0}^{L}\frac{x^{n}}{n!} 
&=&\ln\Big(e^{x}-\sum_{n=L+1}^{\infty}\frac{x^{n}}{n!}\Big) 
\nonumber\\
&=&x + \ln\Big(1-e^{-x}\sum_{n=L+1}^{\infty}\frac{x^{n}}{n!}\Big)
\nonumber\\
&\approx& x - \frac{x^{L+1}}{(L+1)!} \ .
\end{eqnarray}
It can be shown that the higher order terms omitted in Eq.\ (\ref{13})
give only small corrections to the expressions obtained at low temperatures, 
because of the small parameter $|\log{T}|^{-1}\ll1$.
Now the action $S(u,v)$ reads as
\begin{equation}
\label{14}
S(u,v) \approx S_{0}(u,v) + S_{int}(u,v) \ ,
\end{equation}
where
\begin{mathletters}
\begin{eqnarray}
\label{15}
& &S_{0}(u,v)=\sum_{\bf k}(u^{*}_{\bf k}u_{\bf k}+v^{*}_{\bf k}v_{\bf k}-
e^{-\beta \omega _{\bf k}}v^{*}_{\bf k}u_{\bf k}-u^{*}_{\bf k}v_{\bf k}) \ ,\\
\label{16}
& &S _{int}(u,v)=\sum_{i}\frac{(v^{*}_{i}u_{i})^{L+1}}{(L+1)!} \ .
\end{eqnarray}
\end{mathletters}
The term $S_{int}(u,v)$ is of fundamental importance. Indeed, omitting it 
from Eq.\ (\ref{14}) yields the partition function:
$$
Z_{0} = \int e^{-S_{0}(u,v)}DuDv,
$$
which suffers from infrared divergences in the thermodynamic average for 
the one- and two-dimensional cases:
$$
\langle v^{*}_{i}u_{i}\rangle _{0}
=\frac{1}{N}\sum_{\bf k}\frac{1}{e^{\beta\omega_{\bf k}}-1}=\infty \ .
$$
Thus, it is equivalent to the omission of the projection operator 
from Eq.\ (\ref{2}). 

\begin{figure}
\psfig{file=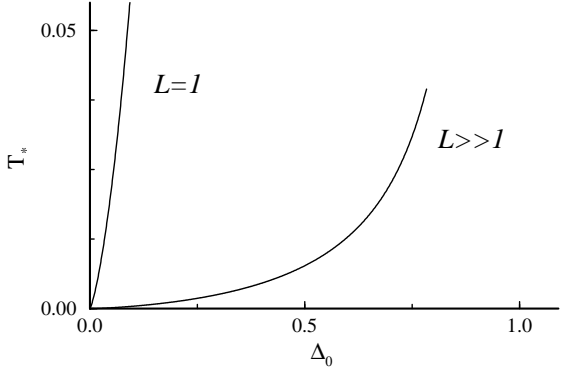,width=3.3in,angle=-90}
\caption{The temperature dependence of $\Delta_0$ for two cases: $L=1$ and $L\gg1$.}
\label{fig2}
\end{figure}

We now linearize the ``action'' Eq. (\ref{14}) as follows:
\begin{equation}
\label{19}
S_{int}(u,v) \rightarrow
\sum_{i}\Big( (L+1)\Delta ^{L}v^{*}_{i}u_{i} 
- L\Delta ^{L+1}\Big) \ 
\end{equation}
and introduce the mean field partition function
\begin{equation}
\label{22}
Z_{mf}(\Delta) = \int e^{-S_{mf}(u,v,\Delta )}DuDv \ ,
\end{equation}
with 
\begin{eqnarray}
\label{20}
S_{mf}(u,v,\Delta) &=& \sum_{\bf k}(u^{*}_{\bf k}u_{\bf k} +
                                    v^{*}_{\bf k}v_{\bf k}-
e^{-\beta \omega _{\bf k}}v^{*}_{\bf k}u_{\bf k}
\nonumber\\
& &{ } - (1 - (L+1)\Delta ^{L})u^{*}_{\bf k}v_{\bf k} - L\Delta ^{L+1}) \ .
\end{eqnarray}
Here we defined the thermodynamic average:
\begin{equation}
\label{21}
\Delta = \langle v^{*}_{i}u_{i}\rangle _{mf} \ .
\end{equation}
Thus, in the mean field approximation, we have the quadratic mean field 
``action'' $S_{mf}(u,v,\Delta)$ instead of nonlinear the ``action'' $S(u,v)$.
The parameter $\Delta$ will be self-consistently obtained below, 
in accordance with Eq.\ (\ref{21}). The simplicity of the mean field 
``action'' (\ref {20}) allows us to perform functional integration in the 
partition function to yield:
\begin{equation}
\label{23}
Z_{mf}(\Delta) = \prod _{\bf k}\frac{e^{L\Delta ^{L+1}}}
                   {1-e^{-\beta \omega _{\bf k}} (1-(L+1)\Delta ^{L})} =
e^{-\beta F_{mf}(\Delta)} \ .
\end{equation}
Here
\begin{eqnarray}
\label{24}
F_{mf}(\Delta) &=& \frac{1}{\beta}\sum_{\bf k}\Big(-L\Delta ^{L+1} 
\nonumber\\
& &{ } +\ln\big(1-e^{-\beta \omega _{\bf k}} (1-(L+1)\Delta ^{L})\big)\Big) \ .
\end{eqnarray}
From the condition:
\begin{equation}
\label{25}
\frac{\partial F_{mf}(\Delta)}{\partial \Delta}\Big| _{\Delta _{0}} = 0 \ ,
\end{equation}
we obtain the self-consistent equation for $\Delta _{0}$~:
\begin{equation}
\label{26}
\Delta _{0} = \frac{1}{N}\sum_{\bf k}\frac{1}
{e^{\beta \omega _{\bf k}}-(1-(L+1)\Delta ^{L}_{0})} \ .
\end{equation}
In the two-dimensional case at low temperature (and for $\Delta_0\ll1$), it reduces to
\begin{equation}
\Delta_{0}=-\frac{T}{2\pi J}\frac{\ln{((L+1)\Delta^{L}_{0})}}{1-(L+1)\Delta_0^L}
\simeq-\frac{T}{2\pi J}\ln{((L+1)\Delta ^{L}_{0})} \ .
\end{equation}
The temperature dependence for $\Delta_0$ is presented in Fig.\ \ref{fig2} 
for two limiting cases: $L=1$ and $L\gg1$ (here $L=500$).
The thermodynamic average $\Delta_0$ vanishes at $T=0$ and rapidly grows as 
temperature increases. When $T_*=T/2\pi J\ll1$,
\begin{equation}
\Delta_0\simeq -T_*L\ln{(T_*)} \ .
\end{equation}

Now, it is obvious that we can exclude the projecting operator from 
Eq.\ (\ref{2}), but simultaneously we must introduce the chemical potential
\begin{equation}
\mu = T(L+1)\Delta ^{L}_{0} = T e^{-2\pi J\Delta _{0}/T}
\end{equation}
in Eq.\ (\ref{1}). This is precisely the Takahashi's assumption
that results in very good agreement with the Bethe-anzatz for an 
one-dimensional ferromagnet at low temperature. Note, $\mu$ decreases with $L$
and vanishes in the absence of the kinematical interaction, $L=\infty$
(i.e., non-interacting bosons).
From Eqs. (23), (24), we obtain the low-temperature asymptotics:
\begin{equation}
\mu\simeq 2\pi J\, T_*^{L+1}\ .       
\end{equation}

To conclude, we have shown that a kinematical interaction in 2D magnon 
systems can be compensated via introducing a ``fictitious'' chemical potential
into their effective dispersion law, and proposed a systematic, 
self-consistent procedure of calculating its low-temperature dependence 
in the one-loop approximation.

We would like to acknowledge P.H. Diamond for his interest and critical suggestions.

\end{document}